\documentclass{ws-ijmpa}

\begin{document}

\markboth{G. Hao, L. Li, C. F. Qiao} {$D_s$ Asymmetry}

%%%%%%%%%%%%%%%%%%%%% Publisher's Area please ignore %%%%%%%%%%%%%%%
%
\catchline{}{}{}{}{}
%
%%%%%%%%%%%%%%%%%%%%%%%%%%%%%%%%%%%%%%%%%%%%%%%%%%%%%%%%%%%%%%%%%%%%

\title{$D_s$ Asymmetry in Photoproduction%\footnote{For the title, try not to use more than
%3 lines. Typeset the title in 10 pt roman, uppercase and
%boldface.}
}

\author{\footnotesize Gang
Hao\footnote{ Email:hao$_-$gang@mails.gucas.ac.cn}}

\address{Dept. of Physics, Graduate School of the Chinese
Academy of Sciences,\\
YuQuan Road 19A, Beijing 100049, China }

\author{Lin Li\footnote{Email:lilin@gucas.ac.cn}}

\address{Dept. of Physics, Graduate School of the Chinese
Academy of Sciences,\\
YuQuan Road 19A, Beijing 100049, China }

\author{Cong-Feng Qiao\footnote{Email:cfqiao@gucas.ac.cn}}

\address{CCAST(World Lab.), P.O. Box 8730, Beijing 100080, China\\
Dept. of Physics, Graduate School of the Chinese
Academy of Sciences,\\
YuQuan Road 19A, Beijing 100049, China }

\maketitle

%\pub{Received (Day Month Year)}{Revised (Day Month Year)}

\begin{abstract}
By adopting two models of  strange and antistrange quark
distributions inside nucleon, the light-cone meson-baryon
fluctuation model and the effective chiral quark model, we
calculate the $D_s^+ - D_s^-$ asymmetry in photoproduction in the
framework of heavy-quark recombination mechanism. We find that the
effect of asymmetry of strange sea to the $D_s$
 asymmetry is considerable and depending on the different models. Therefore,
 we expect that with the further study in electroproduction, e.g.
 at HERA and
 CEBAF, the experimental measurements on the $D_s^+ - D_s^-$ asymmetry may
impose a strong restriction on the strange-antistrange
distribution asymmetry models.
\end{abstract}

\section{Introduction}    %) A SECTION HEADING

      The production of charmed hadron at high-energy colliders has
   been the subject of considerable interest in recent years.
   Particularly, it is still controversial that fixed-target hadro- and
   photoproduction experiments have observed large production
   asymmetries between the charmed and anticharmed mesons.
   \cite{ex1,ex2,ex3,ex4,cha1,cha2,cha3}.In fact, the experimental
   asymmetries of
   charm hadron production are very large compared with the predictions of
    perturbative QCD(pQCD)\cite{nlo1,nlo2,nlo3,nlo4}.

   The
   charm hadrons are heavy enough that the cross section for their
   production can be factorized into a short-distance part, the cross
   section of the production of the $c\bar{c}$ pair, and a
   long-distance parameter, the  nonperturbative fragmentation
   function. In the skeleton of pQCD, the charm-anticharm
   asymmetry comes only from the next-to-leading order(NLO), or
   higher, corrections and is relatively small. Though many attempts
   try to resolve this problem\cite{mod1,mod2}, they all depend
   on the unknown distribution of partons in the remnant of the
   target nucleon or photon after the collider to a great extent.

    Different from these, the heavy-quark recombination mechanism
    proposed by Braaten et al.\cite{Braaten} can give a more
    quantitative and simple explanation to charm meson
    asymmetries.In their works\cite{Braatenc}, Braaten et al.
    consider that the light quark($u,d$) that participates in the
    hard-scattering process may recombine with a heavy
    quark($c,\bar{c}$) in the final state, provided the light quark has
    momentum of $O(\Lambda_{QCD})$ in rest frame of heavy quark.
    And the product of the recombination hadronize into the final
    state heavy meson($D^+,D^-$) while the recoiling heavy
    quark($\bar{c},c$) fragments to  $D^-_s,D^+_s$ meson.
     Because the
    strange sea of nucleon is symmetry in their assumption, the
     $D_s^+ - D_s^-$
    asymmetry is due to the excess of $u$ and $d$ over
    $\bar{u}$ and $\bar{d}$ in proton. The asymmetry of $D_s$
    meson has the opposite sign as that of D meson and is
    relatively small. However, we note that the striking
    strange sea asymmetry in the momentum distribution of the
    nucleon has been proposed\cite{sea1,sea2}. Based on this
    idea, we study the $D_s$ asymmetry in
    photoproduction with the heavy-quark recombination mechanism
    and expect the experimental measurements on this asymmetry
    will impose a restriction on the strange sea distribution in
    nucleon.

\section{The $D_s$ production mechanism}
    The cross section of $D_s$ meson photoproduction takes the
    form in the heavy-quark recombination mechanism:
    \begin{equation}
   d \sigma[\gamma+N \rightarrow D_s +X]=f_{q/N} \otimes \sum d
   \hat{\sigma}[\gamma+\bar{s} \rightarrow (c \bar{s} )^n
   +\bar{c}]\rho [(c \bar{s} )^n \rightarrow D_s],
   \end{equation}
   Here $(\bar{c} s)^n$ indicates the $s$
  has very small momentum in the $\bar{c}$ rest frame, and $n$
  is the color and angular momentum quantum numbers of $(\bar{c} s)$
  recombination. $d \hat{\sigma}[\gamma+ \bar{s} \rightarrow (c
  \bar{s})^n +\bar{c}]$ is the short-distance partonic
  subprocess. The factor $\rho[(c \bar{s})^n \rightarrow D_s]$ is
  the probability of the $(\bar{c} s)^n$ state to hadronize into a
  the $D_s$ final state. The $D_s$ meson is produced by two
  schemes,
   \begin{equation}
  (a)~~ d \hat{\sigma}[\gamma+ \bar{s} \rightarrow (c \bar{s})^n
  +\bar{c}]\rho [(c \bar{s})^n \rightarrow D_s]\;, \label{recom}
  \end{equation}
 \begin{equation}
 (b)~~ d \hat{\sigma}[\gamma+ q \rightarrow (\bar{c} q)^n
 +c]\sum_{\bar D} \rho [(\bar c q)^n \to \bar{D}] \otimes D_{c
 \rightarrow D_s}\; , \label{recom-frag}
 \end{equation}
 In process (a), the $(c \bar{s})^n$ recombines into the $D_s$
 meson directly; in process (b), $(\bar{c} q)^n$ recombines into
 the $D_s^-$, $D^-$ or $\bar{D}^0$ meson, and the recoiling $c$
 quark fragments to the $D_s^+$ meson. Compare with Braaten's
 work, we take into account the asymmetry effects due to the
 process (a).

    To get the total cross section of $D_s$ production,
 it is important to quote the asymmetric distribution of strange
 sea. Here we adopt two distribution models of $s$ quark,i.e.,
 the light-cone meson-baryon
fluctuation model and the effective chiral quark
model\cite{sea1,sea2}. In the first model, the asymmetry of
strange sea comes from the intermediate $K^+\Lambda$ configuration
of the incident nucleon. The second one is based on the effective
chiral theory, in which the asymmetry is caused by both the
constituent quark distribution and the quark splitting function.
More detailed description can be found in Ref.\cite{sea1,sea2}.

With the resultant production cross section, we can get the $D_s^+
- D_s^-$ Asymmetry in Photoproduction from the definition below,
\begin{eqnarray}
\alpha[D_s]&=& \frac{\sigma_{D_s^+}-
\sigma_{D_s^-}}{\sigma_{D_s^+}+ \sigma_{D_s^-}}\; .
\end{eqnarray}

Our prediction of $D_s^+ - D_s^-$ asymmetry, in comparison with
Bratten's result, are shown in Fig.1.

\begin{figure}
\centerline{\psfig{file=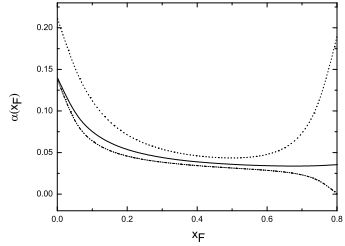,width=5cm}} \vspace*{8pt}
\caption{The asymmetry $\alpha[D_s]$ versus $x_F$. The dotted and
dash-dotted lines correspond to the results from the light-cone
meson-baryon fluctuation model and the effective chiral quark
model, respectively. The solid line from the
 Braatenc's result.}
\end{figure}

\section{Summary}
The detailed results of calculation and the selection of
 the relevant parameters can be find in Ref.\cite{plb}. As can be
 seen from Fig.1, the production asymmetry of $D_s$ by adopting the
 light-cone meson-baryon fluctuation model is about 1.2 times
 larger than Ref.\cite{Braatenc}. While the result of the
 effective chiral quark model is only about $80\%$of the
 Braaten's.In summary, the different models of the distribution
 of the strange sea give the obviously divergent curves of
 the asymmetry versus $x_F$.Though there is still some
 uncertainties, such as the breakdown of $SU(3)$ symmetry,the
 large $N_c$ limit and the NLO correction in pQCD, our results
 show that the effect of asymmetry of strange sea to the $D_s$
 asymmetry is considerable and depending on the different models.
 Moreover, it is noted that the $D_s$ asymmetry experimental data
 are still limited and preliminary. We expect the more precise and
 new experimental data of $D_s$ asymmetry can impose a clear
 restriction on the strange and antistrange quark distribution.

 Data
 from several recent experiments, including the HAPPEX,
   the SAMPLE experiment at
   MIT-Bates and the A4 experiment at the Mainz
 Laboratory in Germany suggest that the  strange quarks may contribute
  a positive value to
 the proton's magnetic moment\cite{1,2,3}. These stimulating experimental
  results
 are beginning to shed further light on the distribution of the strange
 quark. The next promising issues to be studied include confronting
  our theory
 predictions\cite{will} on the $D_s$ asymmetry with the corresponding
 experimental data in electroproduction. We also expect that the
 $D_s$ production asymmetry can be
 observed at HERA and CEBAF, which may give further information on the
 distribution of the strange quark inside the nucleon.

%This section should come before the References. Dedications and funding
%information may also be included here.

%\appendix

\end{document}